# Crystal structure of dense pseudo-cubic boron allotrope, pc-B$_{52}$, by powder X-ray diffraction

Oleksandr O. Kurakevych [a] and Vladimir L. Solozhenko [b,*]

[a] IMPMC, Université P & M Curie, 75015 Paris, France
[b] LSPM–CNRS, Université Paris Nord, 93430 Villetaneuse, France

During past years, a number of reports have been published on synthesis of tetragonal allotrope of boron, t-B$_{52}$ phase. However, no unambiguous characterization of the crystal structure has been performed to the present time, while remarkable variation of the *a/c* lattice-parameter ratio raises strong doubts about its uniqueness. Here the Rietveld refinement of the crystal structure of the high pressure – high temperature boron phase synthesized by a direct solid-state transformation of rhombohedral β-B$_{106}$ at 20 GPa and 2500 K has been reported for the first time. Although this boron allotrope belongs to the t-B$_{52}$ type, its structure can be considered as pseudo-cubic with the *a/c* ratio of √2.

* vladimir.solozhenko@univ-paris13.fr

First report on tetragonal phase of boron, t-B$_{52}$ has appeared in the literature as early as in 1943 [1,2]. However, the result was not reproduced later, and in the following decades t-B$_{52}$ was believed to be an experimental artifact related to the impurities [3-6]. Only quite recently, after strong scientific attention given to boron and boron-rich solids during last years [7-10], a number of reports on the preparation of pure boron allotrope t-B$_{52}$ have appeared. It can be obtained by the crystallization of amorphous boron during a slow heating in the H$_2$/Ar atmosphere [11], by phase transformations of crystalline boron under extreme *p-T* conditions [11], and even by high pressure – high temperature (HPHT) decomposition of boranes [12].

The crystal structure(s) of t-B$_{52}$ is (are) often believed to be close to original Hoard's model [1,11,12], although it was predicted by *ab initio* calculations to be extremely unstable [10], and alternative tetragonal (or even orthorhombic) structures are more preferable [13]. Single-crystal diffraction studies can hardly be performed in the near future because of the difficulties of a crystal growth control during solid-phase transformations under required conditions, while the analysis of a limited number of powder X-ray diffraction (XRD) patterns [11,12] indicates that the lattice parameters vary in a wide range without any correlation with synthesis conditions (Fig. 1a).

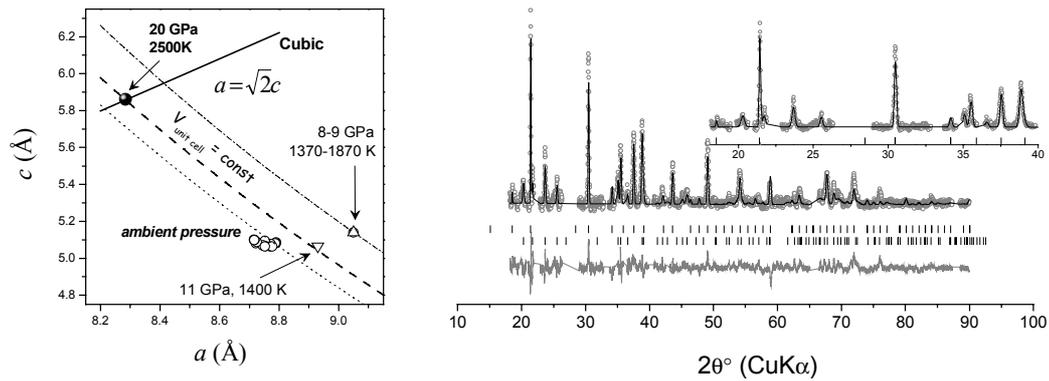

**Figure 1.** (*a*) Lattice parameters of various phases with structure related to t-$B_{52}$. Solid circle (●) represents pc-$B_{52}$ phase synthesized by us, open circles (○) show the lattice parameters of phases obtained at ambient pressure ($B_{50}C_2$ and $B_{50}B_2$) (see [11] and references therein), triangles correspond to t-$B_{52}$ phases synthesized at high pressures from β-$B_{106}$ (▽) [11] and borane (△) [12]. (*b*) Rietveld refinement (GSAS) of pc-$B_{52}$ crystal structure (present work).

New high pressure – high temperature boron phase (HPHT boron) has been synthesized from highly crystalline rhombohedral β-$B_{106}$ (99.995 at %) at 20 GPa and 2500 K using a large-volume multianvil two-stage system. The HPHT assembly has been described elsewhere [10]. To isolate the sample from the elements of a high-pressure cell, capsules of pyrolytic boron nitride, which do not react with crystalline boron at such temperatures [14,15], have been used. Samples were gradually compressed to 20 GPa at room temperature, heated at 2500 K for 5-10 min and then rapidly (~200 K/s) quenched by switching off the electric power and slowly decompressed. New HPHT boron phase (usually in the form of a composite with γ-$B_{28}$) was recovered as well-sintered cylinders (1-2 mm in diameter, 1-1.5 mm in height) of a black material.

The recovered samples were studied by powder X-ray diffraction using G3000 TEXT (Inel) diffractometer in a Bragg-Brentano geometry employing Cu K$α_1$ radiation. In order to fit experimental powder diffraction pattern (Fig. 1b) with GSAS software, we tested both t-$B_{192}$ [16] and $B_{50}C_2$ [6] as starting unit cells. We have chosen the first one because HPHT boron phase was believed to be of the t-$B_{192}$ family [10,17]. The use of the second model was inspired by recent discoveries of t-$B_{52}$ phase(s) at HPHT conditions [11,12]; but instead of the simplified Hoard's model [1], a more complicated $B_{50}C_2$ unit cell was employed. There are two reasons for that: (1) $B_{50}C_2$ was previously studied in detail on single-crystal samples, and (2) carbon has the closest covalent radius to boron as compared to other elements forming t-$B_{52}$-type compounds. The $B_{50}C_2$ starting model allowed us to obtain the best Rietveld fit and the most reliable structure for our HPHT boron phase. Quite high w$R$p and $R$p values (Table) are due to the presence of γ-$B_{28}$, preferred crystallite orientation typical for boron solids recovered from HPHT conditions [18], and peak asymmetry/shifts [18] due to the accumulation of stacking faults during the transformation [11,19].

The lattice parameters of our HPHT boron are $a = 8.2937(10)$ Å and $c = 5.8636(8)$ Å (tetragonal syngony), which is remarkably different from those for t-$B_{52}$ in Refs. 1-3,6,11,12,14 (see Fig. 1a). In fact, such difference explains difficulties in the recognition of a structural similarity between HPHT boron and t-$B_{52}$. The $a/c$ ratio is $\sqrt{2}$ up to the 0.016 % accuracy (just like ratio between the side of square and a half of its diagonal). The powder XRD pattern also well fits the cubic structure with $a = 8.294$ Å (space group of NaCl, i.e. $Fm\bar{3}m$). Thus, in order to distinguish our dense HPHT boron phase from other members of the t-$B_{52}$ family, we will name it "pseudo-cubic $B_{52}$", or simply pc-$B_{52}$.

To the best of our knowledge, pc-$B_{52}$ has been obtained at the highest temperature among all known phases of the t-$B_{52}$ family (see Fig. 1a). At such temperatures atomic diffusion becomes sufficiently fast to result in the formation of a thermodynamically stable allotrope. Its density is 2.528 g cm$^{-3}$, just slightly below the value for dense superhard γ-$B_{28}$ [10,20]. All this allows one to suggest that pc-$B_{52}$ should be as hard and incompressible as γ-$B_{28}$ [21-23].

**Table 1.** Results of Rietveld refinement of crystal structure of pc-$B_{52}$.

| Unit cell | Atomic parameters | | | | | Quality of refinement |
|---|---|---|---|---|---|---|
| | Name | x | y | z | Occupancy | |
| P42/nnm (setting 2) | B1 | 0.537(2) | -0.181(2) | 0.664(2) | 1.0 | 43 reflections |
| | B2 | 0.500(2) | -0.174(2) | 0.327(2) | 1.0 | |
| | B3 | 0.393(2) | -0.107(2) | 0.595(3) | 1.0 | |
| $a = 8.2937(10)$ Å | B4 | 0.490(2) | -0.010(2) | 0.830(3) | 1.0 | w$R$p = 0.20 |
| $c = 5.8636(8)$ Å | B5 | 0.25 | -0.25 | 0.25 | 1.0 | $R$p = 0.15 |
| | B6 | 0.904(3) | -0.25 | 0.25 | 0.1 | |
| ρ = 2.528 g cm$^{-3}$ | B7 | 0.25 | 0.25 | 0.331(5) | 0.5 | $\chi^2$ = 12.5 for 17 variables |
| $V_{at}$ = 4.27 cm$^3$ mol$^{-1}$ | B8 | 0.25 | -0.25 | 0.75 | 1.0 | |

Refined atomic coordinates of pc-$B_{52}$ (see the Table) allowed us to calculate the lengths of B–B bonds and angles between them. Some unrealistic interatomic lengths (below 1.4 Å), which formally appear due to the partial unoccupancy of some crystallographic places, are indicative of a temperature-induced structural disorder, similar to $B_{50}C_2$ [6] and β-$B_{106}$ [24]. In contrast, "low-temperature phases" α-$B_{12}$ and γ-$B_{28}$ do not show such disorder [10,25]. One can easily see the similarity in the distribution of bonding lengths and angles for $B_{50}C_2$ and pc-$B_{52}$ (Fig. 2a and 2b); while in the case of "low-temperature" γ-$B_{28}$ that is stable at the same pressures as pc-$B_{52}$, the distributions are much narrower.

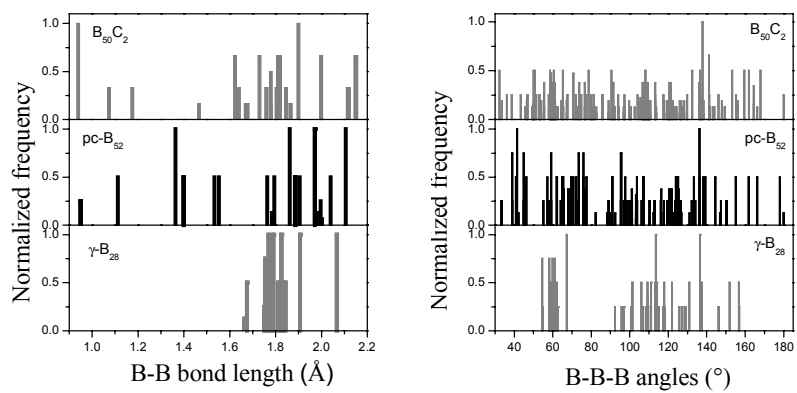

**Figure 2.** (*a*) Bond length distributions in pc-$B_{52}$, $B_{50}C_2$ [6] and $\gamma$-$B_{28}$ [10]. (*b*) Interbond angle distributions in pc-$B_{52}$, $B_{50}C_2$ [6] and $\gamma$-$B_{28}$ [10]. Normalized frequency represents the number of bonds *N* having a given length (or a given angle) per unit cell (divided by the maximal *N*).

Finally, we have resolved the crystal structure of HPHT boron, which for a long time was supposed to be structurally similar to t-$B_{192}$ [10,17]. Now this phase can be considered as pseudo-cubic pc-$B_{52}$ that belongs to the t-$B_{52}$ family of boron allotrope(s) and compounds [6,11,26,27]. Comparison of pc-$B_{52}$ and $\gamma$-$B_{28}$ shows the close values of key properties responsible for hardness (density, coordination number, bonding type, etc. [28]); thus, pc-$B_{52}$ is expected to be superhard and low-compressible [8], in contrast to other t-$B_{52}$ phases.

**Acknowledgements.** This work was financially supported by the Agence Nationale de la Recherche (grant ANR-2011-BS08-018).